# Prédiction de la surface foliaire du tournesol au stade végétatif par analyse d'image et application pour l'estimation de paramètres de réponses au stress hydrique chez des variétés en post-inscription

Pierre Casadebaig, Nicolas Blanchet et Nicolas Langlade
06/10/2021

## Résumé

La mesure automatique des réponses développementale et physiologique du tournesol au stress hydrique représente un enjeu appliqué pour une meilleure connaissance des variétés disponibles pour les cultivateurs mais aussi fondamentale pour identifier les bases biologiques, génétiques et moléculaires de la réponse des plantes à leur environnement.

Nous avons mis en place sur la plateforme de phénotypage haut-débit Heliaphen de INRAE Toulouse deux expérimentations chacune avec 8 variétés (96 plantes) et acquis des images par une barrière lumineuse de façon quotidienne de plantes soumises ou non à un stress hydrique. En parallèle, nous avons mesuré les surfaces foliaires de ces plantes tous les deux jours de manière manuelle pendant la durée du stress d'environ dix jours. Les images ont été analysées pour extraire des caractéristiques morphologiques des plantes segmentées et différents modèles ont été évaluées pour estimer les surfaces foliaires totales de plantes à l'aide de ces données.

Un modèle linéaire avec lissage a posteriori permet d'estimer la surface foliaire totale avec une erreur quadratique relative de 11% et une efficience de 93%. Les surfaces foliaires estimées de façon classique ou avec le modèle développé ont permis de calculer les réponses de l'expansion foliaire et de la transpiration (LER et TR) utilisés dans le modèles de culture SUNFLO pour 8 variétés de tournesol étudiées. Des coefficients de corrélation de 0.61 et 0.81 pour LER et TR respectivement valide l'usage de l'estimation des surfaces foliaires par imagerie. On note toutefois des estimations de valeurs plus faibles pour LER que par méthode manuelle sur Heliaphen mais plus proches globalement de la méthode manuelle sur des plantes élevées en serresuggérant potentiellement une surestimation de la sensibilité au stress.

On peut conclure que les estimations des paramètres LE et TR peuvent être utilisées pour des simulations. Le faible coût de cette méthode (comparée aux mesures manuelles), la possibilité de paralléliser et répéter les mesures sur la plateforme Heliaphen et de bénéficier de la gestion des données de la plateforme Heliaphen, constituent des améliorations majeures pour la valorisation du modèle SUNFLO et la caractérisation de la sensibilité à la sécheresse des variétés cultivées.

## Objectifs

Les objectifs de cette collaboration étaient de :

- Compléter l'évaluation post-inscription de deux séries de variétés commercialisées fournies par Terres Inovia en les phénotypant sur la plateforme Heliaphen pour leurs paramètres de réponse au déficit hydrique utiles au modèle SUNFLO,
- Évaluer la précision de prédiction de la surface foliaire totale au cours du temps par les algorithmes d'analyse d'images aujourd'hui disponibles,
- Estimer et comparer les seuils de réponse d'expansion foliaire (LER) et de transpiration (TR) des variétés de tournesol en étude en utilisant les estimations de surface foliaire par méthode manuelle et par analyse d'images.

## Matériel et méthodes

L'expérimentation 18HP10 a été réalisée du 2018-09-14 au 2018-09-26 et 19HP10 du 2019-05-29 au 2019-06-17. Dans les deux expérimentations, 8 génotypes ont été étudiés avec deux conduites hydriques : contrôle (irrigué en ciblant la capacité du pot) et stressé (arrêt de l'irrigation et dessèchement progressif). Chaque expérimentation est basée sur 96 pots (2 traitements, 4 répétitions pour les plantes irriguées et 8 répétitions pour les plantes stressées et 8 génotypes).

Pour 18HP10, l'évolution de la surface de la plante est mesurée manuellement (mesure des surfaces des feuilles, tous les deux jours) et automatiquement (acquisition quotidiennes de variables caractéristiques de l'architecture). Pour 19HP10, l'évolution de la surface est mesurée automatiquement, avec deux dates manuelles utilisées pour la validation. Pour chaque expérimentation, un rapport additionnel détaille la méthode d'analyse et les résultats.

## Résultats

### Prédiction de la surface foliaire par analyse d'image.

Un capteur de type barrière lumineuse disposé sur le robot de phénotypage scanne la plante lors des opérations de pesées et d'arrosage, et produit une image (profil de la plante). L'analyse de ces images permet d'extraire 70 variables qui sont utilisées comme prédicteurs de la surface de la plante. Dans une première étape, nous avons confronté trois approches de prédiction, pour comparer la performance d'un modèle de régression (Generalized Linear Model, GAM) et d'un modèle d'apprentissage (Multi-layer Perceptron). Nous avons ensuite comparé la précision d'un modèle ajusté pour toutes les variétés et conditions à des modèles spécifiques aux traitements ou variétés.

#### Choix de l'approche de prédiction

Une première analyse indique que la surface projetée (area_sens), la surface enveloppe (hull_area), la surface de la boîte englobante (bounding_rectancle), et la hauteur de la plante sont linéairement corrélées à la surface mesurée (figure 1). Ces quatre variables sont utilisées dans le modèle de régression, pour éviter une étape automatique de sélection de variable.

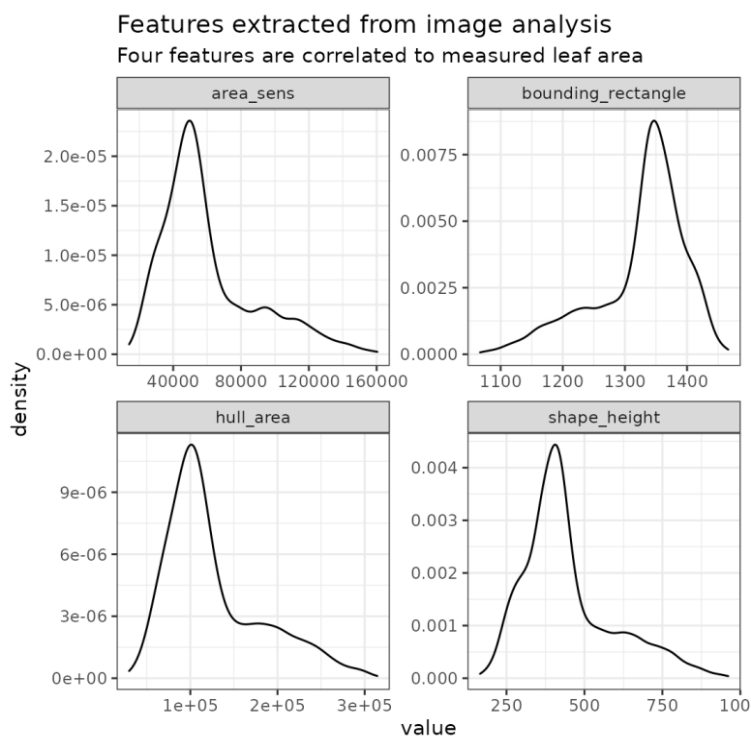

*Figure 1. Distribution des 4 variables les plus corrélées à la surface de la plante mesurée.*

La comparaison entre les approches de prédiction indique que le modèle de régression a des performances plus faibles, mais relativement comparable à une approche d'apprentissage (figure 2, panneaux 1 et 2).

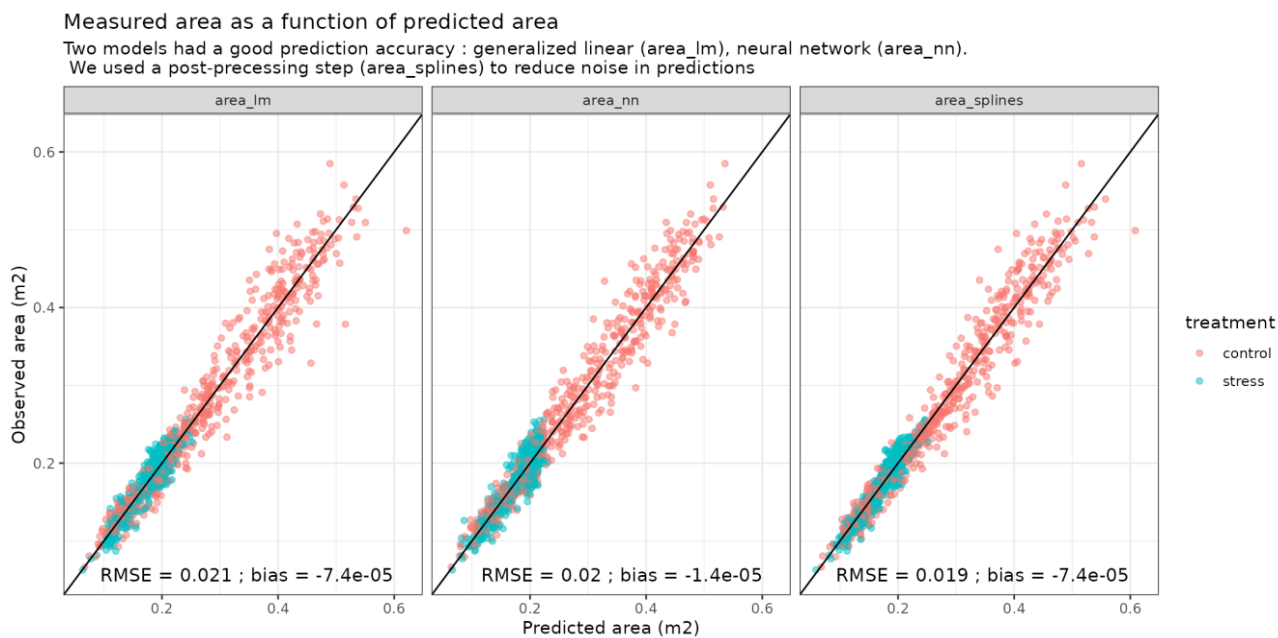

*Figure 2. Comparaison des approches de prédictions utilisées. Modèle linéaire généralisé (lm, gauche), réseau de neurones (nn, centre), post-precessing avec lissage (splines, droite)*

Dans les deux cas, les prédictions de surface ne sont pas strictement croissantes avec le temps, ce qui invalide l'utilisation de ces prédictions pour l'analyse écophysiologique ultérieure, basée sur le calcul du taux d'accroissement journalier. Ce problème a été résolu avec une étape de post-traitement, basée sur un lissage des prédictions de surfaces brutes (figure 2, droite). Cette étape permet d'obtenir des taux d'accroissement strictement positifs ou nuls (figure 3). La précision de l'approche prédiction et post-traitement est au final légèrement meilleure que les modèles de prédiction bruts (figure 2, panneau 3).

*Table 1. Comparaison des approches de prédictions utilisées.*

*n : effectif, rmse : erreur quadratique (mm²), rmse_rel : erreur quadratique normalisée (%), biais (mm²) et efficience.*

| method | n | rmse | rmse_rel | bias | efficiency |
|---|---|---|---|---|---|
| area_lm | 1238 | 20895.47 | 0.09 | -73.84 | 0.95 |
| area_nn | 1185 | 20362.21 | 0.09 | -13.59 | 0.96 |
| area_splines | 1238 | 18907.59 | 0.08 | -73.84 | 0.96 |

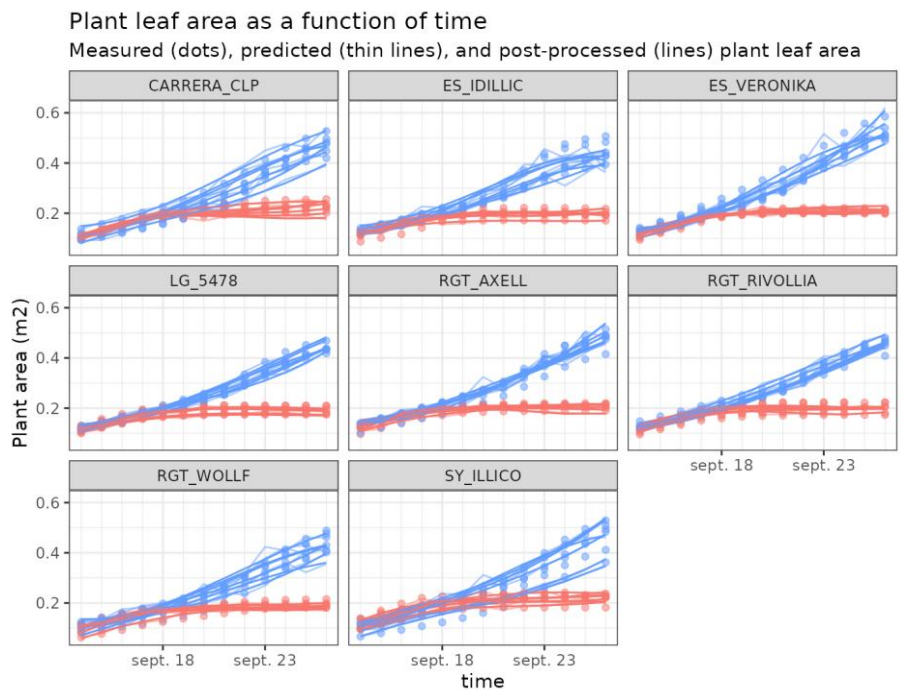

*Figure 3. Effet de la procédure de post-traitement sur les courbes de croissance en surface.*

**Choix du paramétrage du modèle**

Un gain de prédiction important n'est visible que dans le cas du développement d'un modèle par variété. Cette stratégie est difficile sur le plan opérationnel, car elle nécessite de calibrer le modèle pour les nouvelles variétés avant leur phénotypage sur la plateforme Heliaphen. Un modèle global (ou un modèle par traitement d'irrigation) présente une erreur de prédiction plus élevée, mais acceptable. C'est cette option qui est utilisée pour prédire la surface foliaire de l'expérimentation 19HP10.

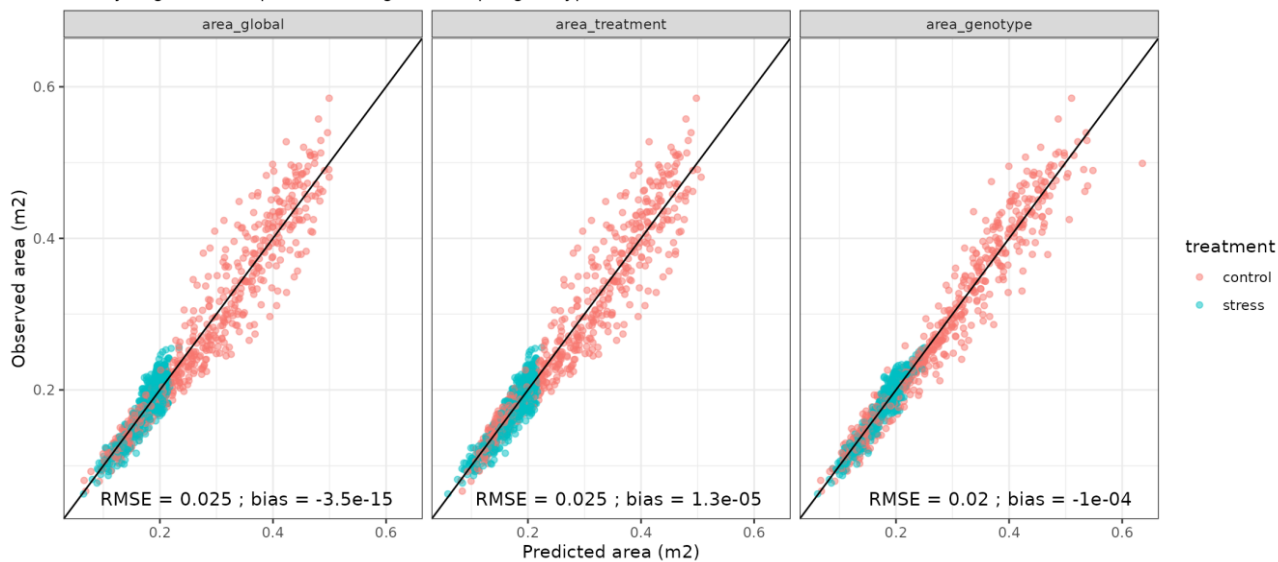

*Figure 4. Effet de l'échelle de modélisation sur la qualité de prédiction.*

*Table 2. Comparaison des échelles de modélisation.*

*n : effectif, rmse : erreur quadratique ($mm^2$), rmse_rel : erreur quadratique normalisée (%), biais ($mm^2$) et efficience.*

| method | n | rmse | rmse_rel | bias | efficiency |
|---|---|---|---|---|---|
| area_genotype | 1238 | 19686.69 | 0.09 | -101.76 | 0.96 |
| area_global | 1238 | 25080.81 | 0.11 | 0.00 | 0.93 |
| area_treatment | 1238 | 25291.15 | 0.11 | 13.22 | 0.93 |

**Phénotypage des caractéristiques de réponse au déficit hydrique pour l'évaluation variétale**

L'analyse de la relation entre les processus physiologiques et le déficit hydrique permet de déterminer une réponse propre à chaque variété (table 3). La précision des deux expérimentations a été suffisante pour identifier des valeurs de paramètres différents de la valeur moyenne des variétés.

*Table 3. Synthèse du paramétrage variétal pour la tolérance au déficit hydrique des processus d'expansion foliaire (LE) et de transpiration de la plante (TR).*

*Une valeur plus faible de paramètre (estimate) indique une variété qui commence à adapter son expansion ou transpiration à un fort niveau de contrainte hydrique (FTSW).*

| process | genotype | estimate | se | rmse |
|---|---|---|---|---|
| Expansion foliaire LE | ES_ISIDA | -2.85 | 0.18 | 0.18 |
| | RGT_SITTINGBULL | -3.22 | 0.14 | 0.11 |
| | MAS_87A | -3.33 | 0.13 | 0.10 |
| | LG_5679 | -3.40 | 0.20 | 0.15 |
| | SY_CELESTO | -3.40 | 0.18 | 0.15 |
| | RGT_GLLOSS | -3.46 | 0.19 | 0.14 |
| | MAS_98K | -3.68 | 0.17 | 0.12 |
| | SY_CHRONOS | -3.77 | 0.27 | 0.17 |
| | RGT_AXELL | -2.65 | 0.22 | 0.20 |
| | LG_5478 | -2.69 | 0.26 | 0.23 |
| | ES_VERONIKA | -2.82 | 0.27 | 0.23 |
| | ES_IDILLIC | -2.90 | 0.30 | 0.24 |
| | RGT_RIVOLLIA | -2.93 | 0.27 | 0.23 |
| | RGT_WOLLF | -3.47 | 0.30 | 0.20 |
| | SY_ILLICO | -4.00 | 0.47 | 0.25 |
| | CARRERA_CLP | -4.55 | 0.49 | 0.23 |
| Transpiration TR | RGT_AXELL | -5.01 | 0.49 | 0.24 |
| | ES_IDILLIC | -5.12 | 0.47 | 0.22 |
| | CARRERA_CLP | -5.40 | 0.41 | 0.17 |
| | RGT_RIVOLLIA | -5.45 | 0.44 | 0.19 |
| | SY_ILLICO | -5.49 | 0.48 | 0.19 |
| | ES_VERONIKA | -5.63 | 0.53 | 0.23 |
| | RGT_WOLLF | -6.64 | 0.71 | 0.23 |
| | ES_ISIDA | -6.02 | 0.39 | 0.16 |
| | SY_CHRONOS | -6.45 | 0.55 | 0.17 |
| | LG_5679 | -7.59 | 0.54 | 0.16 |
| | RGT_SITTINGBULL | -8.01 | 0.48 | 0.13 |
| | MAS_87A | -8.34 | 0.50 | 0.14 |
| | RGT_GLLOSS | -8.40 | 0.58 | 0.16 |
| | MAS_98K | -8.68 | 0.56 | 0.14 |
| | SY_CELESTO | -8.83 | 0.60 | 0.16 |
| | LG_5478 | -4.99 | 0.40 | 0.19 |

**Impact de la méthode de mesure de surface foliaire sur les caractéristiques variétales.**

Sur l'expérimentation 18HP10, les paramètres de réponse des variétés ont été calculés selon les deux méthodes de mesure de la surface (manuelle et automatique). La valeur de ces paramètres est affectée par la méthode de mesure de la surface, mais nous observons une corrélation modérée (R = 0.61 et 0.81 respectivement). Le biais important visible pour le processus d'expansion indique que l'utilisation de la méthode de mesure automatique de la surface semble générer des paramètres avec des valeurs faibles, c'est-à-dire une réponse physiologique seulement pour des déficits hydriques élevés.

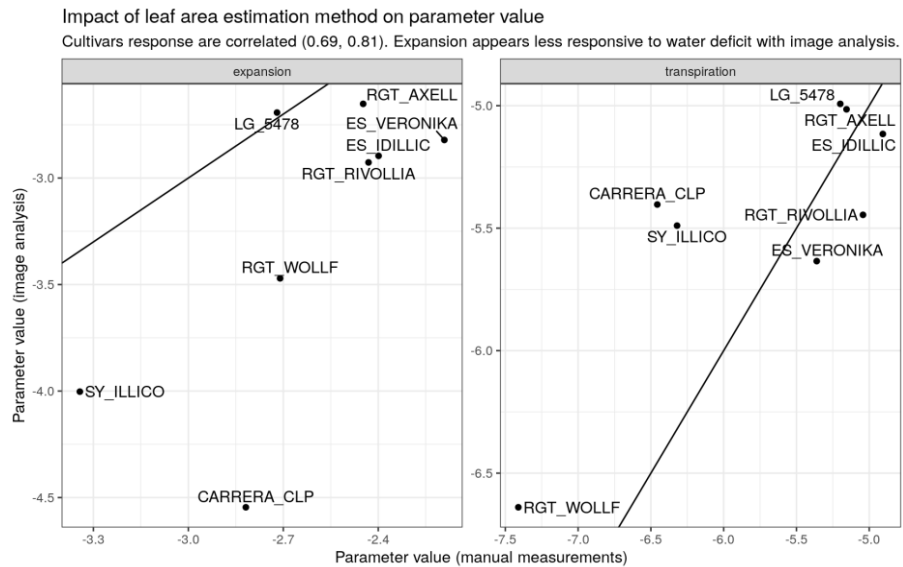

*Figure 5: Corrélation entre les paramètres d'expansion foliaire (LE, gauche) et de transpiration (TR, droite) entre la méthode d'estimation manuelle en abscisses et par analyse d'images en ordonnées*

## Conclusion et perspectives

Le projet EPhECTS a permis de développer un méthode d'estimation des paramètres du modèle de culture SUNFLO caractérisant les sensibilités de l'expansion foliaire et de la transpiration sur 18 variétés cultivées de tournesol lors de deux expérimentations en 2018 et 2019.

Dans un premier temps, trois méthodes d'estimation de la surface foliaire à partir d'images acquises automatiquement par la plateforme Phenotoul-Heliaphen et traitées par le logiciel IPSOPhen (développé par INRAE) ont été évaluées. En utilisant 1238 surfaces acquises manuellement et automatiquement, les trois méthodes donnent des résultats assez proches, la méthode avec lissage a posteriori étant la plus performante sur les données disponibles a été retenue car elle permet en outre de produire des croissances foliaires nulles ou positives.

L'utilisation d'information a priori sur le dispositif expérimental a mis en évidence que seul la connaissance du génotype améliore la qualité de prédiction et que la connaissance du régime hydrique n'a pas d'influence significative sur la qualité de prédiction. Nous pouvons donc utiliser un modèle général d'estimation de la surface foliaire quel que soit le traitement hydrique et le génotype.

Les données d'estimation de surface foliaires ont été exploitées avec les pesées automatiques du robot pour estimer les seuils de réponses de l'expansion foliaire et de la transpiration vis-à-vis du stress hydrique utilisés dans SUNFLO. Les paramètres LE et TR ont été estimés pour 18 variétés, ils sont distribués entre -2.85 et -4.55 (LE) et -4.99 et -8.83 (TR). La gamme des variétés EPhECTS est identique à celle précédemment mesurée manuellement dans (Gosseau et al., 2019) pour LE et dans le haut de la gamme pour TR.

La comparaison des paramètres obtenus avec des mesures manuelles de surface folaiires ou de façon automatique par analyse d'images a ensuite pu être réalisée sur une partie du dispositif : l'expérimentation de 20218. Premièrement, les erreurs relatives pour les estimations des paramètres sont entre 0.10 et 0.25 (LE) et entre 0.13 et 0.24 (TR) et sont donc supérieures aux erreurs obtenues avec des mesures manuelles et présentées dans (Gosseau et al., 2019). De plus, un biais sur les estimations de LE est observé : des valeurs plus faibles sont obtenues avec la méthode par analyse d'image pour LE. On peut noter qu'au contraire nous avions obtenu des valeurs plus élevées en passant de mesures manuelles en serre vers des mesures manuelles sur Heliaphen. La méthode par analyse d'images produit donc des estimation dans une gamme de valeurs plus proche avec la méthode originale d'estimation de ces paramètres SUNFLO.

Pour le paramètre de sensibilité de la transpiration, l'interprétation de la corrélation est risquée car celle-ci est portée par une variété qui est insensible au stress hydrique (RGT WOLLF). Cependant, il n'y a pas de biais systématique apparent.

Globalement, le projet EPhECTS a permis de développer une méthode rapide, peu onéreuse et haut-débit d'évaluation des paramètres de sensibilité à la sécheresse du tournesol.

Les estimations utilisant le traitement d'image sont corrélées à celles obtenues par méthode de référence mais présentent des erreurs supérieures.

On peut conclure que les estimations des paramètres LE et TR peuvent être utilisées pour des simulations. Le faible coût de cette méthode (comparée aux mesures manuelles), la possibilité de paralléliser et répéter les mesures sur la plateforme Heliaphen et de bénéficier de la gestion des données de la plateforme Heliaphen, constituent des améliorations majeures pour la valorisation du modèle SUNFLO et la caractérisation de la sensibilité à la sécheresse des variétés cultivées.

### Perspectives

Trois perspectives au projet peuvent être identifiées :

1. la comparaison des simulations avec SUNFLO et des données de rendement réelles en utilisant des valeurs estimées par analyse d'images et de référence obtnues dans EPhECTS

2. l'amélioration des modèles d'estimation de surface foliaire avec des jeux de données plus importants

3. le durcissement des pipelines d'analyse utilisés dans EPhECTS pour de futurs travaux nécessitant l'estimation de ces paramètres, par exemple pour la caractérisation post-inscription des variétés du catalogue français ou celle de variétés ou génotypes utilisés dans des projets de recherche

## Bibliographie